\documentclass[aps,twocolumn,showpacs,showkeys,groupedaddress,amsmath,amssymb,showpacs,showkeys]{revtex4}
\usepackage{amssymb}
\usepackage{aas_macros}
\usepackage{graphicx}

\begin{document}

\title{Equations of motion, initial and boundary conditions for GRB}
\author{C.L. Bianco, R. Ruffini, G.V. Vereshchagin, S.-S. Xue}
\affiliation{I.C.R.A. - International Center for Relativistic Astrophysics, \\
University of Rome "La Sapienza", Physics Department,
P.le A. Moro 5, 00185 Rome, Italy.}

\date{29 of September 2005}

\begin{abstract}
We compare and contrast the different approaches to the optically thick adiabatic phase of GRB all the way to the transparency. Special attention is given to the role of the rate equation to be self consistently solved with the relativistic hydrodynamic equations. The works of Shemi and Piran \cite{1990ApJ...365L..55S}, Piran, Shemi and Narayan \cite{1993MNRAS.263..861P}, M\'{e}sz\'{a}ros, Laguna and Rees  \cite{1993ApJ...415..181M} and Ruffini, Salmonson, Wilson and Xue \cite{1999A&A...350..334R},\cite{2000A&A...359..855R}  are compared and contrasted. The role of the baryonic loading in these three treatments is pointed out. Constraints on initial conditions for the fireball produced by electro-magnetic black hole are obtained. 
\end{abstract}



\maketitle

\section{Introduction}

The phenomenon of gamma ray bursts (GRBs) attracts a lot of attention in the modern relativistic astrophysics. In spite of revolutionary progress in observational technics and instruments over the past decades, GRBs remain misterious events since their discovery. Although the nature of the energy which powers GRBs remains in general uncertain, there exists a conventional model which describes the phenomenon since the moment when the huge energy is released in a compact spatial region. It is referred to as the \emph{cosmic fireball} model \cite{1999PhR...314..575P}.

The qualitative study of cosmic fireballs was given in the pioneering paper by Cavallo and Rees \cite{1978MNRAS.183..359C}. Here we remind briefly their main results. The authors considered sudden energy release in a compact spherical region of space in the form of photons. If the initial average photon energy in units of electron rest mass energy in the fireball is larger than unity,
\begin{eqnarray}
    \epsilon \equiv \frac{E_0}{m_{e}c^{2}}>1,
\label{epsilon}
\end{eqnarray}
the fireball is opaque due to large number of electron-positron pairs produced. These pairs in turn produce new photons, sharing their energy, so the systems cools at the expense of growing number of particles. It becomes transparent only when $\epsilon \leq 1$, and pairs annihilate. However, this behavior changes if admixture of a plasma is also present in the fireball. This admixture is conveniently parametrized by

\begin{equation}
\eta =\frac{E_{0}}{Mc^{2}},
\end{equation}

\bigskip where $E_{0},M,c$ are initial total energy of the fireball, total mass of the plasma and the speed of light. So initial optical depth of the fireball is given by

\begin{equation}
\tau_0 =\sigma _{T}R_0(n^0_{\pm }+n^0_{gas}),
\end{equation}

where $\sigma _{T}$ is the Thomson cross section, $R_0$ is initial radius of the fireball and $n^0_{\pm},n^0_{gas}$ are initial number densities of electon-positron pairs and ionized gas (which we will call simply plasma in the following) respectively.

The plasma is assumed to be in the form of ionized hydrogen and electons, totally neutral, and it could be enough plasma to make the fireball optically thick even \emph{after} the energy decreases below the pair production threshold. Then the fireball starts to expand adiabatically and its energy diminishes as inverse radius until the condition $\tau\simeq 1$ is fulfilled. Once the optical depth reaches unity, the fireball becomes transparent and photons stream away. This is a basic qualitative model, which was developed in great details in subsequent years. Unfortunately it was abondoned later in favor of other models implementing for instance the notion of beaming. In particular, great attention was paid to the interaction of expanding relativistic baryons left from the fireball with interstellar medium (ISM) surrounding the fireball.

The aim of the paper is to give a critical review of existing models for isotropic relativistic fireballs, compare and contrast these models. The paper is organized as follows. In the next section we give the basic equations and describe approximations involved into description of the fireballs. Then we present our model which differs from models in the literature as it describes the dynamics of the fireball taking into account the \emph{rate equations} for electron-positron pairs. Then we compare various models for the fireball. In the final section we discuss initial conditions for the fireball and impose corresponding constraints arising from our model. Conclusions follow.

\section{Energy-momentum principle}

The basis of description for relativistic fireballs is the energy-momentum principle. It allows to obtain relativistic hydrodynamical equations, or equations of motion for the fireball, energy and momentum conservation equations which are used extensively to describe interaction of relativistic baryons of the fireball with the interstellar gas, and boundary conditions which are used to understand shock waves propagation in the decelerating baryons and in the outer medium. Consider energy-momentum conservation in the most general form\footnote{Greek indices denote four-dimensional components and run from 0 to 3 while Latin indices run from 1 to 3. The general relativistic effects are neglected, which is a good approximation, but we left the general definition of the energy-momentum conservation to take into account arbitrary coordinate system.}:

\begin{eqnarray}
    T^{\mu\nu}{}_{;\nu}=\frac{\partial (\sqrt{-g}\,T^{\mu\nu})}{\partial x^\nu}+\sqrt{-g}\,\Gamma^\mu_{\nu\lambda} T^{\nu\lambda}=0,
\label{ce}
\end{eqnarray}
where $\Gamma^\mu_{\nu\lambda}$ are Cristoffel symbols and $g$ is determinant of the metric tensor.
Integrating over the whole three-dimensional volume we obtain
\begin{eqnarray}
    \int_V T^{\mu\nu}{}_{;\nu} dV=0.
\label{cev}
\end{eqnarray}
Integrating over the whole four-dimensional volume and applying divergence theorem we get \cite{1948PhRv...74..328T}
\begin{eqnarray}
    \int_t\int_V T^{\mu\nu}{}_{;\nu} dV dt=\oint_V T^{\mu\nu} \lambda_\nu dV=0,
\label{ceom}
\end{eqnarray}
where $\lambda_\alpha$ are covariant components of the outward drawn normal to the three-dimensional hypersurface (volume $V$).

Define the momentum four-vector $P^\mu$:
\begin{eqnarray}
    P^\mu=\int_V T^{0\mu} dV.
\label{mom}
\end{eqnarray}
From (\ref{mom}) and (\ref{cev}) in Minkowski metric (when $\Gamma^\mu_{\nu\lambda}=0$) we see that
\begin{eqnarray}
    \frac{dP^0}{dt}=\int_V \frac{\partial T^{00}}{\partial t} dV = -\int_V \frac{\partial T^{i0}}{\partial x^i} dV = -\oint_S T^{i0}dS_i, \\
    \frac{dP^j}{dt}=\int_V \frac{\partial T^{0j}}{\partial t} dV = -\int_V \frac{\partial T^{ij}}{\partial x^i} dV = -\oint_S T^{ij}dS_i, \\
\end{eqnarray}
so if the energy and momentum fluxes through the surface $S$ bounding considered volume $V$ are absent the energy and momentum are constants during system evolution. Supposing this is the case we arrive to the conservation of energy and momentum:
\begin{eqnarray}
    P^\mu=\mathrm{const}.
\label{consv}
\end{eqnarray}
This equation is important to describe interaction of the baryons left from the fireball with the interstellar gas. Assume the energy-momentum tensor in the form of the ideal fluid
\begin{eqnarray}
    T^{\mu\nu}=p\,g^{\mu\nu}+\omega\,U^\mu U^\nu,
\end{eqnarray}
where $\omega=\epsilon+p$ is proper entalpy, $p$ is proper pressure and $\epsilon$ is proper energy densities. Now suppose spherical symmetry\footnote{The only nonvanishing components of the energy-momentum tensor are $T^{00},\,T^{01,},\,T^{10},\,T^{11},\,T^{22},\,T^{33}.$}, which is usually done for fireballs description. Using spherical coordinates with the interval
\begin{eqnarray}
    ds^2=-dt^2+dr^2+r^2 d\theta^2+r^2\sin^2\theta d\varphi^2,
\end{eqnarray}
we rewrite (\ref{ce}):
\begin{eqnarray}
    \frac{\partial T^{00}}{\partial t}+\frac{1}{r^2}\frac{\partial}{\partial r}\left(r^2 T^{01}\right)=0, \\
    \frac{\partial T^{10}}{\partial t}+\frac{1}{r^2}\frac{\partial}{\partial r}\left(r^2 T^{11}\right)-r\left(T^{33}+T^{44}\sin^2\theta \right)=0,
\end{eqnarray}
arriving to equations of motion for relativistic fireballs \cite{1993MNRAS.263..861P,1993ApJ...415..181M,1999A&A...350..334R,1976PhFl...19.1130B,1995PhRvD..52.4380B}:
\begin{eqnarray}
\label{conseq1}
    \frac{\partial(\gamma^2\omega)}{\partial t}-\frac{\partial p}{\partial t}+\frac{1}{r^2}\frac{\partial}{\partial r}\left(r^2\gamma^2 u \omega\right)=0, \\
    \frac{\partial(\gamma^2 u\omega)}{\partial t}+\frac{1}{r^2}\frac{\partial}{\partial r}\left[r^2(\gamma^2-1) \omega\right]+\frac{\partial p}{\partial r}=0,
\label{conseq2}
\end{eqnarray}
where the four-velocity and the relativistic gamma factor are defined as follows:
\begin{eqnarray}
    U^\mu=(\gamma,\gamma u,0,0), \quad \quad \gamma\equiv(1-u^2)^{-1/2},
\end{eqnarray}
the radial velocity $u$ is measured in units of speed of light
$u=v/c$.

Now suppose that there is a discontinuity on the fluid flow. Suppose the three-dimensional volume is a spherical shell and choose the coordinate system where the discontinuity is at rest so that in (\ref{ceom}) for normal vectors to the discontinuity hypersurface $\lambda_\alpha$ we have
\begin{eqnarray}
    \lambda_\alpha \lambda^\alpha=1, \quad \quad \lambda_0=0.
\end{eqnarray}
Let the radius of the shell $R_s$ be very large and shell thickness $\Delta$ be very small. With $R_s\rightarrow\infty$ and $\Delta\rightarrow 0$ from (\ref{ceom}) we arrive to
\begin{eqnarray}
    \left[ T^{\alpha i} \right]=0,
\end{eqnarray}
where the brackets mean that the quantity inside is the same on both sides of the discontinuity surface. This equation together with continuity conditon for particle density flux $[n U^i]=0$ was used by Taub \cite{1948PhRv...74..328T} to obtain relativistic Rankine-Hugoniot equations, describing shock waves dynamics. Relativistic shocks are supposed to appear during collision of the baryonic material left from the fireball with the ISM \cite{1976PhFl...19.1130B}. The origin of the afterglow could be connected to the conversion of kinetic energy into radiative energy in these shocks \cite{1992MNRAS.258P..41R,1992ApJ...395L..83N,1994ApJ...422..248K,1999PhR...314..575P}. However, our scenario differs from that, namely we suppose that fully radiative condition during this interaction is satisfied. Our model allows to explain the afterglow phenomenon without consideration of shocks \cite{2003AIPC..668...16R} as sources of radiation.

\section{Quasi-analytic model of GRBs}
\label{qam}

The first detailed models for relativistic fireballs were suggested in the beginning of nineties \cite{1990ApJ...365L..55S,1993MNRAS.263..861P,1993ApJ...415..181M}. Independent calculations performed in \cite{1999A&A...350..334R} and \cite{2000A&A...359..855R} give precise understanding and we describe our approach first, mentioning the main differences with the existing literature.

First of all, the source of energy, being obscure in previous models, is supposed to be the energy extraction process from electromagnetic black hole (EMBH) \cite{2003AIPC..668...16R}. The second difference is that initially not photons but pairs are created by overcritical electric field of the EMBH and these pairs produce photons later. This plasma, referred to as pair-electro-magnetic (PEM) pulse expands initially into the vacuum surrounding the black hole reaching very soon relativistic velocities. Then collision with the baryonic remnant of the EMBH takes place and the PEM pulse becomes pair-electro-magnetic-baryonic (PEMB) pulse (see \cite{2003AIPC..668...16R} for details). It was shown that the final gamma factor does not depend on the distance to the baryonic remnant and parameters of the black hole. The only crucial parameters are again the energy of the dyadosphere, or simply $E_0$, and baryonic admixture:

\begin{equation}
B=\frac{Mc^{2}}{E_{0}}=\eta ^{-1}.
\end{equation}

The exact model is based on numerical integration of relativistic energy-momentum conservation equations (\ref{conseq1},\ref{conseq2}) together with the baryonic number conservation equation\footnote{Instead of (\ref{conseq1}) the projection on the flow line $U_\mu(T^{\mu\nu}{})_{;\nu}=0$ is used in \cite{1999A&A...350..334R} and \cite{2000A&A...359..855R}.}
\begin{eqnarray}
(n_B U^\mu)_{;\mu}=0.
\label{conseq3}
\end{eqnarray}
However, the most important distinct point from all previous models is that the \emph{rate equation} for electron-positron pairs is added to the model and integrated simultaneously in order to reach self-consistency.

Here we concentrate on the simple quasi-analytical treatment presented in \cite{1999A&A...350..334R,2000A&A...359..855R} (see also \cite{2003AIPC..668...16R}). The PEMB pulse is supposed to contain finite number of shells each with flat density profile. The dynamics is determined by the following set of equations:

\begin{eqnarray}
\frac{n_{B}^{0}}{n_{B}} &=&\frac{V}{V_{0}}=\frac{\mathcal{V}}{\mathcal{V}_{0}%
}\frac{\gamma }{\gamma _{0}},  \label{eqn_rr_1} \\
\frac{\tilde\epsilon _{0}}{\tilde\epsilon } &=&\left( \frac{V}{V_{0}}\right) ^{\tilde\Gamma}=\left(
\frac{\mathcal{V}}{\mathcal{V}_{0}}\right) ^{\tilde\Gamma}\left( \frac{\gamma }{\gamma
_{0}}\right) ^{\tilde\Gamma},  \label{eqn_rr_2} \\
\frac{\gamma }{\gamma _{0}} &=&\sqrt{\frac{(\tilde\epsilon _{0}+\rho _{B}^{0})%
\mathcal{V}_{0}}{(\tilde\epsilon +\rho _{B})\mathcal{V}},}  \label{eqn_rr_3}
\end{eqnarray}

where $\tilde\Gamma\simeq \frac{4}{3}$ is a thermal index giving the pressure $p=(\tilde\Gamma-1)\tilde\epsilon$, $\tilde\epsilon$ is a proper internal energy density $\tilde\epsilon=\rho-\rho_B$, $\rho _{B}=n_{B}m_{p}c^{2}$ is baryon mass density in comoving frame, $V$ and $\mathcal{V}$ are the proper volume in the comoving frame and the volume in the coordinate frame: $V=\gamma (r)\mathcal{V}$. Subscript "0" denotes initial values, and all quantities are assumed being averaged over finite distibution of shells with constant width and density profiles. All components such as photons, electrons, positrons and plasma ions give contribution to energy density and pressure. This set of equations is equivalent to (\ref{eqn_sp_adi}) and (\ref{eqn_sp_gamma}) (see below). The next step is to take into account the rate equation:

\begin{equation}
\frac{\partial }{\partial t}N_{e^{\pm }}=-N_{e^{\pm }}\frac{1}{\mathcal{V}}
\frac{\partial \mathcal{V}}{\partial t}+\overline{\sigma \mathfrak{v}}\frac{1
}{\gamma ^{2}}(N_{e^{\pm }}^{2}(T)-N_{e^{\pm }}^{2}),  \label{eqn_rr_4}
\end{equation}

where $\sigma $ is the mean pair annihilation-creation cross section, $\mathfrak{v}$ is the thermal velocity of $e^{\pm }$-pairs. The coordinate number density of $e^{\pm }$-pairs in equilibrium is
$N_{e^{\pm}}(T)=\gamma n_{e^{\pm }}(T)$ and the coordinate number density of $e^{\pm }$-pairs is
$N_{e^{\pm }}=\gamma n_{e^{\pm }}$. For $T>m_{e}c^{2}$ we have $n_{e^{\pm}}(T)=n_{\gamma }(T)$, i.e. the number densities of pairs and photons are equal. The pair number densities are given by appropriate Fermi integrals with zero chemical potential, at the equilibrium temperature $T$.

For an infinitesimal expansion of the coordinate volume from $\mathcal{V}_{0} $ to $\mathcal{V}$ in the coordinate time interval $t-t_{0}$ one can discretize the last differential equation for numerical computations.

The most importants outcomes from analysis performed in \cite{2000A&A...359..855R} are the following:

\begin{itemize}
\item[-] the appropriate model for geometry of expanding fireball (PEM-pulse) is given by the constant width approximation (this conclusion is achieved by comparing results obtained using (\ref{conseq1},\ref{conseq2}) and simplified treatment described above),

\item[-] there is a bound on parameter $B$ which comes from violation of constant width approximation, $B\leq 10^{-2}$ ($\eta \geq 10^{2}$).
\end{itemize}

The last conclusion is crucial since it shows that there is a critical loading of baryons, when their presence produce a turbulence in the outflow from the fireball, its motion becomes very complicated and the fireball evolution does not lead in general to the GRB.

Exactly because of this reason, the optically thick fireball never reaches such large radius as $r_b=r_0\eta^2$ (discussed in \cite{1993ApJ...415..181M}, see section (\ref{lmr})) since to do this, the baryonic fraction should overcome the critical value $B_c=10^{-2}$. For larger values of $B_c$ the theory reviewed here does not apply. This means in particular, that all conclusions in \cite{1993ApJ...415..181M} obtained for $r>r_b$ are invalid. In fact, for $B<B_c$ the gamma factor even does not reach saturation (see sec. \ref{lmr}).

\begin{figure}
    \centering
        \includegraphics[width=3in]{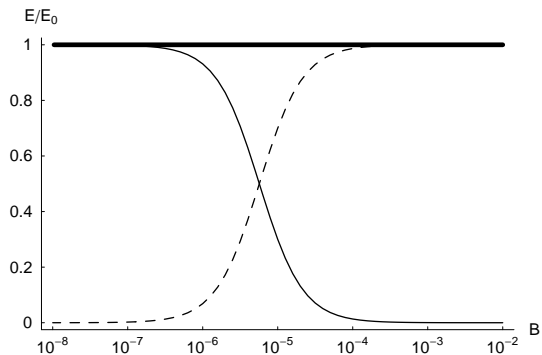}
    \caption{Relative energy release in the form of photons emitted at transparency point $E_{\gamma}/E_0$ (solid line) and kinetic energy of the plasma $E_k/E_0$ (dashed line) of the baryons in terms of initial energy of the fireball depending on parameter $B=\eta^{-1}$ obtained on the basis of quasi-analytic model. Thick line denotes the total energy of the system in terms of initial energy $E_0$.}
    \label{diag}
\end{figure}

\begin{figure}
    \centering
        \includegraphics[width=3in]{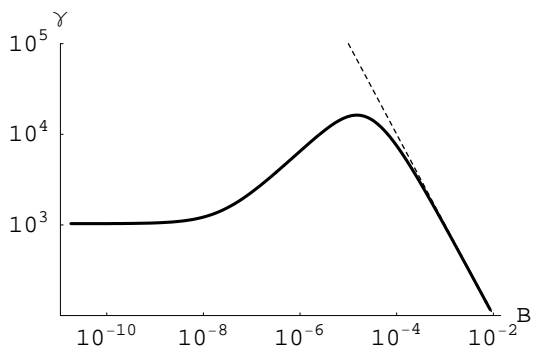}
    \caption{Relativistic gamma factor of the fireball when it reaches trasparency depending on the value of parameter $B$. Dashed line gives asymptotic value $\gamma=B^{-1}$.}
    \label{gammab}
\end{figure}

The fundamental results coming from this model are the diagrams presented at fig. \ref{diag} and \ref{gammab}. The first one shows basically which portion of initial energy is emitted in the form of gamma rays $E_\gamma$ when the fireball reaches transparency condition $\tau\simeq 1$ and how much energy gets converted into the kinetic form of the baryons $E_k$ left after pairs annihilation and photons escape.

The second one gives the value of gamma factor at the moment when the system reaches transparency.

The energy conservation holds, namely
\begin{eqnarray}
    E_0=E_\gamma+E_k,
\label{enRcon}
\end{eqnarray}
Clearly, if the baryons abundance is low most energy is emitted when the fireball gets transparent. It is remarkable that almost all initial energy is converted into kinetic energy of baryons already in the region of validity of constant thickness approximation $B<10^{-2}$, so the region $10^{-8}<B<10^{-2}$ is the most interesting from this point of view.

\section{Shemi and Piran model}
\label{sp}

In this section we discuss the model, proposed by Shemi and Piran \cite{1990ApJ...365L..55S}. This quantitative model gives rather good general picture of relativistic fireballs.

Shemi and Piran found that the temperature at which the fireball becomes optically thin is determined as

\begin{equation}
\mathcal{T}_{esc}=\min (\mathcal{T}_{g},\mathcal{T}_{p}),
\end{equation}

where $\mathcal{T}_{g}$ and $\mathcal{T}_{p}$\ is the temperature when it reaches transparency with respect to gas (plasma) or pairs:

\begin{eqnarray}
\mathcal{T}_{g}^{2} & \simeq & \frac{45}{8\pi^{3}}\frac{m_{p}}{m_{e}}\frac{1}{\alpha^{2}
g_{0}^{\frac{1}{3}}}\frac{1}{\mathcal{T}_{0}^{2}\mathcal{R}0}\eta,
\label{eqn_sp_Tg} \\
\mathcal{T}_{p} & \simeq & 0.032,
\end{eqnarray}

where $m_{p},m_{e},$ are proton and electron masses, $g_{0}=\frac{11}{4},$ $\alpha =\frac{1}{137},$ dimensionless temperature $\mathcal{T}$ and radius $\mathcal{R}$ of the fireball are measured in units of $\frac{m_{e}c^{2}}{k}$ and $\lambda _{e}\equiv \frac{\hbar}{m_{e}c}$ correspondingly, and the subscript "0" denotes initial values. The temperature at transparency point in the case when plasma admixture is unimportant
is nearly a constant for a range of parameters of interest and it nearly equals

\begin{equation}
T_{p}=15\,\,\mathrm{keV}.
\label{eqn_Tp}
\end{equation}

Adiabatic expansion of the fireball implies:

\begin{equation}
\frac{\mathcal{E}}{\mathcal{E}_{0}}=\frac{\mathcal{T}}{\mathcal{T}_{0}}=\frac{\mathcal{R}_0}{\mathcal{R}}  \label{eqn_sp_adi},
\end{equation}
where $\mathcal{E}=\frac{E}{m_{e}c^{2}}$ is a radiative energy in terms of electron rest-mass energy. From the energy conservation (\ref{mom}), supposing the fluid to be pressureless
and its energy density profile is constant we have in the coordinate frame:

\begin{eqnarray}
    \int T^{00}d{\cal V}=\gamma^2\rho{\cal
    V}=\gamma\rho V=\gamma E_{tot}=\mathrm{const}.
\end{eqnarray}

Supposing at initial moment $\gamma_0=1$ and remembering that $E_{tot}=E+M c^2$ we arrive to the following fundamental expression of relativistic gamma factor $\gamma $ at transparency point:

\begin{equation}
\gamma =\frac{\mathcal{E}_{0}+\mathcal{M}c^{2}}{\mathcal{E}+\mathcal{M}c^{2}}=
\frac{\eta+1}{(\frac{\mathcal{T}_{esc}}{\mathcal{T}_{0}})\eta +1},
\label{eqn_sp_gamma}
\end{equation}

where $\mathcal{M}=\frac{M}{m_{e}}.$

One can use this relation to get such important characteristics of the GRB as observed temperature and observed energy. In fact, they can be expressed as follows:

\begin{eqnarray}
\mathcal{T}_{obs} & = & \gamma \mathcal{T}_{esc}, \\
\mathcal{E}_{obs} & = & \mathcal{E}_{0}\frac{\mathcal{T}_{obs}}{\mathcal{T}_{esc}}.
\end{eqnarray}

\begin{figure}
    \centering
        \includegraphics[width=3in]{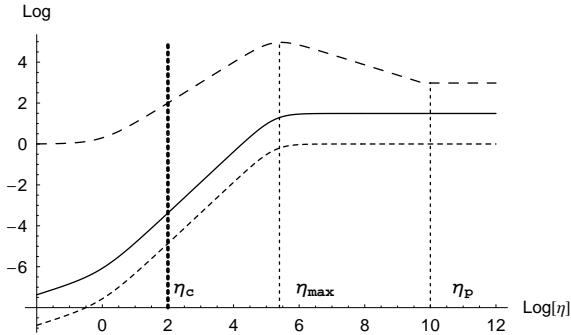}
    \caption{The relativistic gamma factor (upper dashed line), the observed temperature (solid line), and the ratio of observed energy to the initial energy of the fireball (lower dashed line) as a function of $\protect\eta$ (see \cite{1990ApJ...365L..55S}). The values of parameters are the same as in the cited paper.  Thick dashed line denotes the limiting value $\eta_c$. The values of $\eta$ when gamma factor reaches maximum and gets constant are also shown.}
    \label{fig_sp}
\end{figure}

These results are presented at fig. \ref{fig_sp}. In the limit of small $\eta $ we have $\gamma =(1+\eta )$, while, for very large $\eta$ the value of gamma factor at transparency point is $\gamma=\mathcal{T}_{0}/\mathcal{T}_{esc}$, and it has a maximum at intermediate values of $\eta $. We donote by dashed thick line the limiting value of $\eta$ parameter $\eta_c\equiv B_c^{-1}$. For $\eta<\eta_c$ the approximations used to construct the model do not hold. It is clear that because of the presence of bound $\eta_c$ the value $\gamma=\eta$ can be reached only as asymptotic one. In effect, the value $\eta_c$ cuts the region where saturation of the gamma factor happens before the moment when the fireball becomes transparent.

It was found that for relatively large $\eta \geq 10^{5}$ the photons emitted when the fireball becomes transparent carry most of the initial energy. However, since the observed temperature in GRBs is smaller than initial temperature of the fireball, one may suppose that a large part of initial energy is converted to kinetic energy of the plasma.

\section{Shemi, Piran and Narayan model}

Piran, Shemi and Narayan \cite{1993MNRAS.263..861P} present generalization of this model to arbitrary initial density profile of the fireball. These authors performed numerical integrations of coupled energy-momentum relativistic consevation equations (\ref{conseq1},\ref{conseq2}) and baryon number conservation equation (\ref{conseq3}). They were mainly interested in the dependence of the observed temperature, gamma factor and other quantities on the radius of the fireball. Their study results in the number of important conclusions, namely:

\begin{figure}
    \centering
        \includegraphics[width=3in]{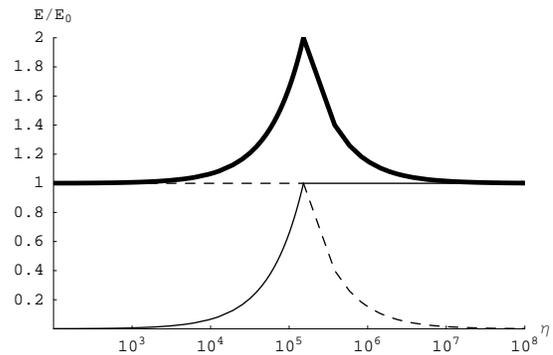}
    \caption{The ratios of radiation and kinetic energy to the initial energy of the fireball predicted by M\'{e}sz\'{a}ros, Laguna and Rees model. Thick line denotes the total energy of the system in terms of initial energy. Energy conservation does not hold.}
    \label{lmr_diag}
\end{figure}
\begin{figure}
    \centering
        \includegraphics[width=3in]{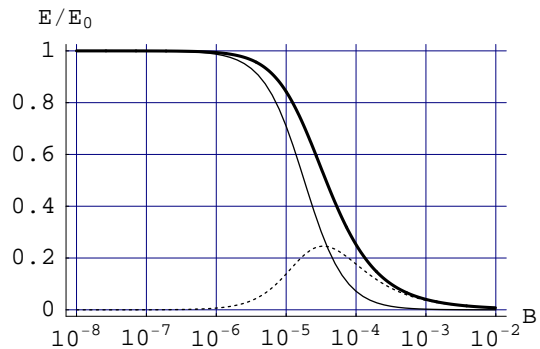}
    \caption{Relative energy release in the form of photons emitted at transparency point $E_{obs}/E_0$ of the GRB in terms of initial energy of the fireball depending on parameter $B=\eta^{-1}$. Thick line represents numerical results and it is the same as in fig. \ref{diag}. Normal line shows results for the analytic model of Shemi and Piran \cite{1990ApJ...365L..55S}. Dashed line shows the difference between exact numerical and approximate analytical results.}
    \label{fig_sp_diag}
\end{figure}

\begin{itemize}

\item[-] the expanding fireball has two basic phases: a radiation dominated phase and a matter-dominated phase. In the former, the gamma factor grows linearly as the radius of the fireball: $\gamma \varpropto r$, while in the latter the gamma factor reaches asymptotic value $\gamma \simeq \eta +1$.

\item[-] the numerical solutions are reproduced with a good accuracy by frozen-pulse approximation, when the pulse width is given by initial radius of the fireball.

\end{itemize}

The last conclusion is important, since the volume $V$ of the fireball can be calculated as

\begin{eqnarray}
V=4\pi R^2 \Delta,
\end{eqnarray}
where $\Delta\simeq R_0$ is the width of the leading shell with consant energy density profile, $R$ is the radius of the fireball.

They also present the following scaling solution:

\begin{eqnarray}
R & = & R_{0}\left( \frac{\gamma _{0}}{\gamma }D^{3}\right)^{1/2}, \\
\frac{1}{D} & \equiv & \frac{\gamma _{0}}{\gamma }+\frac{3\gamma _{0}}{4\gamma \eta }-\frac{3}{4\eta },
\end{eqnarray}

where subscript "0" denotes some initial time when $\gamma \gtrsim$ few, which can be inverted to give $\gamma(R)$.

\section{M\'{e}sz\'{a}ros, Laguna and Rees model}
\label{lmr}

The next step in developing this model was made in \cite{1993ApJ...415..181M}. In order to reconcile the model with observations, these authors proposed a generalization to anisotropic (jet) case. Nevertheless, their analytic results apply to the case of homogeneous isotropic fireballs and we will follow their analytical isotropic model in this
section.

Starting from the same point as Shemi and Piran, consider (\ref{eqn_sp_adi}) and (\ref{eqn_sp_gamma}). The part of the paper, containing analytical results, describes the geometry of the fireball, gamma factor behavior and the final energy balance between radiation and kinetic energy. Magnetic field effects are also considered, but we are not interested in this part here.

Three basic regimes are found in \cite{1993ApJ...415..181M} for evolution of the fireball. In two first regimes there is a correspondence between the analysis in the paper and results of \cite{1993MNRAS.263..861P}, so the constant thickness approximation holds. It is claimed in \cite{1993ApJ...415..181M}, that when the radius of the fireball reaches very large values such as $R_{b}=R_{0}\eta^{2}$ the noticable departure from constant width of the fireball occurs. However, it is important to note, that the fireball becomes transparent much earlier and this effect never becomes important (see section \ref{qam}).

The crucial quantity, presented in the paper is $\Gamma _{m}$ -- the maximum possible bulk Lorentz factor achievable for a given initial radiation energy $E_{0}$ deposited within a given initial radius $R_{0}$:

\begin{eqnarray}
\Gamma _{m} &\equiv &\eta _{m}=\left( \tau _{0}\eta \right) ^{1/3}=\left( \Sigma _{0}\kappa \eta \right) ^{1/3},
\label{eqn_mlr_Gm} \\
\Sigma _{0} &=&\frac{M}{4\pi R_{0}^{2}}, \quad\quad \kappa =\frac{\sigma _{T}}{m_{p}},
\end{eqnarray}

where $\Sigma _{0}$ is initial baryon (plasma) mass surface density.

All subsequent calculations in the paper \cite{1993ApJ...415..181M} involves this quantity. It is evident from (\ref{eqn_mlr_Gm}) that the linear dependence between the gamma factor $\Gamma $ and parameter $\eta $ is assumed. However, this is certainly not true as can be seen from fig. \ref{fig_sp}. We will come back to this point in the following section.

Another important quantity is given in this paper, namely

\begin{equation}
\Gamma _{p}=\frac{T_{0}}{T_{p}}.
\end{equation}

\bigskip This is just the asymptotic behavior of the gamma factor at fig. \ref{fig_sp} for very large $\eta $. Using it, the authors calculate the value of $\eta $ parameter above which the pairs dominated regime occurs:

\begin{equation}
\eta _{p}=\frac{\Gamma _{m}^{3}}{\Gamma _{p}^{2}}.
\end{equation}

This means, \emph{above $\eta _{p}$ the presence of baryons in the fireball is insufficient to keep the fireball opaque after pairs are annihilated} and almost all initial energy deposited in the fireball is emitted immediately.
The estimate of the final radiation to kinetic energy ratio, made in \cite{1993ApJ...415..181M} is incorrect, because kinetic and radiation energies do not sum up to initial energy of the fireball thus violating energy
conservation (\ref{enRcon}). This is illustrated at fig. \ref{lmr_diag}. The correct analytic diagram is presented in fig. \ref{fig_sp_diag} instead.

\section{Approximate results}
\label{approx}

All models for isotropic fireballs are based on the following points:

\begin{enumerate}

\item Flat space-time,

\item Relativistic energy-momentum principle,

\item baryonic number conservation.

\end{enumerate}

Anthough the EMBH model starts with Reissner-Nordstrom geometry, the numerical code is written for the case of flat space-time simply because curved space-time effects becomes insignificant soon after the fireball reachs relativistic expansion velocities. The presence of rate equation in the model \cite{1999A&A...350..334R},\cite{2000A&A...359..855R} has a deep physical ground and its luck in the other treatments means incompleteness of these models. Indeed, the number density of pairs influences the speed of expansion of the fireball. However, in this section we neglect the rate equation and discuss the common points between all considered models.

First of all, let us come back to fig. \ref{fig_sp}. For almost all values of parameter $\eta $ the gamma factor is determined by gas (i.e. plasma or baryons) admixture according to (\ref{eqn_sp_Tg}), consider this case below. For given initial energy and radius this temperature depends only on $\eta $ only, so one can write:

\begin{equation}
\gamma =\frac{\eta +1}{(\frac{\mathcal{T}_{g}}{\mathcal{T}_{0}})\eta +1}=%
\frac{\eta +1}{a \eta ^{\frac{3}{2}}+1},  \label{eqn_an}
\end{equation}

where

\begin{equation}
a =2.1\cdot 10^{3}\mathcal{T}_{0}^{-2}\mathcal{R}_{0}^{-0.5}.
\end{equation}

From this formula we can get immediately the two asymptotic regimes, namely:

\begin{equation}
\gamma =\left\{
\begin{array}{cc}
\eta +1, & \eta <\eta _{\max }, \\
\frac{1}{a \sqrt{\eta }}, & \eta >\eta _{\max }%
\end{array}%
\right. .
\label{eqn_g_as}
\end{equation}

Notice, that the constant $a$ is extremely small number, so that after obtaining precise value of $\eta _{\max }$ by equating to zero the derivative of function (\ref{eqn_an}) one can expand in Taylor series the result and get in the lowest order in $\alpha $, that:

\begin{eqnarray}
\eta _{\max } &\simeq &\left( \frac{2}{a}\right) ^{\frac{2}{3}}-2, \\
\gamma _{\max } &\equiv &\gamma (\eta _{\max })\simeq \frac{1}{3}\left[
1+\left( \frac{2}{a}\right) ^{\frac{2}{3}}\right] .
\end{eqnarray}

In particular, in the case shown in fig. \ref{fig_sp} one has $\eta _{\max }=2.8\cdot 10^{5}$, $\gamma _{\max }=9.3\cdot 10^{4}$ while according to (\ref{eqn_mlr_Gm}) $\Gamma _{m}=\eta _{\max }=1.75\cdot 10^{5}$. Clearly,
our result is much more accurate. Actually, the value $\Gamma_m$ in (\ref{eqn_mlr_Gm}) is obtained from equating asymptotes in (\ref{eqn_g_as}) and there exists the following relation:

\begin{eqnarray}
a=(\tau_0\eta)^{-1/2}.
\end{eqnarray}

Now we can explain why the observed temperature (and consequently the observed energy) does not depend on $\eta$ in the region $\eta_{max}<\eta<\eta_p$. From the second line in (\ref{eqn_g_as}) it follows that the gamma factor in this region behaves as $\gamma\propto\eta^{-1/2}$, while $\mathcal{T}_{esc}\propto\eta^{1/2}$. These two exactly compensate each other leading to independence of the observed quantities on $\eta$ in this region. This remains the same for $\eta>\eta_p$ also, since here $\mathcal{T}_{esc}=\mathcal{T}_{p}=$const and from (\ref{eqn_an}) $\gamma=$const.

\section{Significance of the rate equation}
\label{rate}

The rate equation describes the number densities evolution for electrons and positrons. In analytic models it is supposed that pairs are annihilated instantly when transparency condition is fulfilled. Moreover, the dynamics of expansion is influenced by the electron-positron energy density as can be seen from (\ref{eqn_rr_1})-(\ref{eqn_rr_4}). Therefore, it is important to make clear whether neglect of the rate equation is a crude approximation or not.

Using eq. (\ref{eqn_sp_gamma}) one can obtain analytic dependence of the energy emitted at transparency point on parameter $B$ and we compare it at fig. \ref{fig_sp_diag}.

We also show the difference between numerical results based on integration of eqs. (\ref{eqn_rr_1})-(\ref{eqn_rr_4}) and analytic results from Shemi and Piran model. The values of parameters are: $\mu=10^3$ and $\xi=0.1$ (which correspond to $E_0=2.87\cdot10^{54}\,$ergs and $R_0=1.08\cdot 10^9\,$cm). One can see that the difference peaks at intermediate values of $B$. The crucial deviations however appear for large $B$, where analytical predictions for observed energy are about two orders of magnitude smaller than the numerical ones. This is due to the difference in predictions of the radius of the fireball at transparency moment. In fact, the analytical model overestimates this value at about two orders of magnitude for $B=10^{-2}$. So for large $B$ with correct treatment of pairs dynamics the fireball gets transparent at \emph{earlier} moments comparing to the analytical treatment.

\begin{figure}
	\centering
		\includegraphics[width=3in]{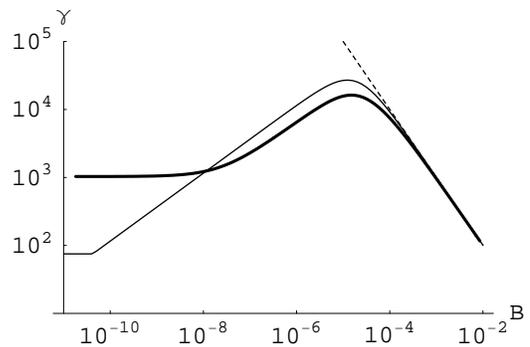}
	\caption{Relativistic gamma factor when transparency is reached. The thick line denotes exact numerical results, the normal line corresponds to analytical estimate from Shemi and Piran model, the dashed line denotes the asymptotic value $\gamma=B^{-1}$.}
	\label{gamma}
\end{figure}

At the same time, the difference between numerical and analytical results for gamma factor is significant for small $B$ as illustrated at fig. \ref{gamma}. While both results coincide for $B>10^{-4}$ there is a constant difference for the range of values $10^{-8}<B<10^{-4}$ and asymptotic constant values for the gamma factor are also different. Besides, this asymptotic behavior takes place for larger values of $B$ in disagreement with analytical expectations. Thus the acceleration of the fireball for small $B$ is larger if one accounts for pairs dynamics.

\begin{table*}
    \centering
\begin{tabular}{||l|l|l|l|l||}\hline
{\small } & {\small Ruffini et. al.} & {\small Shemi, Piran} & {\small Piran, Shemi,
Narayan} & {\small M\'{e}sz\'{a}ros, Laguna, Rees} \\ \hline
{\small conservation:} & & & & \\
{\small energy-momentum,} & {\small yes} & {\small yes} & {\small yes} & {\small yes} \\
{\small baryon number,} & {\small yes} & {\small do not consider} &
{\small yes} & {\small yes} \\
\hline
{\small rate equation} & {\small yes} & {\small no} & {\small no} & {\small %
no} \\
\hline
{\small constant width} & {\small justify} & {\small do not consider%
} & {\small justify} & {\small in part} \\
{\small approximation} & & & & \\
\hline
{\small model for }$\gamma (r)$ & {\small numerical/analytic} &
{\small no} & {\small numerical/analytic} & {\small numerical/analytic} \\
\hline
{\small model for }$\gamma (\eta )$ & {\small numerical} & {\small analytic}
& {\small do not consider} & {\small analytic}%
\\ \hline
\end{tabular}
    \caption{Comparison of different models for fireballs.}
    \label{mod}
\end{table*}

It is clear that the error coming from neglect of the rate equation is significant. This implies that simple analytic model of Shemi and Piran gives only qualititive picture of the fireball evolution and in order to get correct description of the fireball one cannot neglect the rate equation.

Moreover, the difference between exact numerical model \cite{1999A&A...350..334R},\cite{2000A&A...359..855R} and approximate analytical models \cite{1990ApJ...365L..55S} becomes apparent in various physical aspects, namely in predictions of the radius of the shell when it reaches transparency, the gamma factor at transparency and the ratio between the energy released in the form of photons and the one converted into kinetic form. The last point is crucial. It is assumed in the literature, that the whole initial energy of the fireball gets converted into kinetic energy of the shell during adiabatic expansion. Indeed, taking typical value of parameter $B$ as $10^{-3}$ we find that according to Shemi and Piran model we have only $0.2\%$ of initial energy left in the form of photons. However, exact numerical computations \cite{1999A&A...350..334R},\cite{2000A&A...359..855R} give $3.7\%$ for the energy of photons radiated when the fireball reaches transparency, which is a significant value and it cannot be neglected.

To summurize the above discussion, we present the result of this survey in the Table \ref{mod}. It is important to notice again that comparing to simplified analytic treatment, accounting for the rate of change of electron-positron pairs densities gives quantitatively different results on the ratio of kinetic versus photon energies produced in the GRB and the gamma factor at transparency moment, which in turn leads to different afterglow properties. Therefore, although analytical models presented in sections \ref{sp} and \ref{approx} agree and give correct qualitative description of the fireball, one should use numerical approach described in sec. \ref{qam} in order to compare the theory and observations.

\section{Cavallo and Rees diagram and EMBH model}

As already mentioned in the section \ref{qam} the fireball can be produced by EMBH if the electric field strength exceeds some critical value $E_c$. Within the shell surrounding just formed EMBH where $E(r)>E_c$ called \emph{dyadosphere} \cite{1998A&A...338L..87P,2003JKPSP} the overcritical electric field produces electron-positron pairs which in turn produce photons by reactions $\gamma+\gamma\leftrightarrow e^{+}+e^{-}$, quickly thermalize and begin to expand being accelerated by the radiative pressure leading finally to observed GRB.

The fireball can clearly emerge in the EMBH theory only if the dyadosphere radius $r_{dya}$ is larger than the horizon radius $r_{+}$. These quantities are defined as follows \cite{1998A&A...338L..87P,2003JKPSP}:
\begin{eqnarray}
    r_{+}=\frac{GM_\odot}{c^2}\mu\left[1+\sqrt{1-\xi^2}\right]
\end{eqnarray}
and
\begin{eqnarray}
    r_{dya}=\sqrt{\frac{Q_{max}}{4\pi{\cal E}_c}}\sqrt{\mu\xi},
\end{eqnarray}
where $G$ is Newton's gravitational constant, $\mu$ and $\xi$ are parameters of EMBH, $\mu=M/M_\odot$, $\xi=Q/Q_{max}$, $Q_{max}=M\sqrt{G}$, $M_\odot$ is solar mass and the critical electric field strength ${\cal E}_c$ is defined \cite{1931HE,1951PhRv...82..664S} as
\begin{eqnarray}
    {\cal E}_c=\frac{m_e^2 c^3}{\hbar e},
\end{eqnarray}
where $e$ is electron charge and $\hbar$ is Planck's constant.

\begin{figure}
    \centering
        \includegraphics[angle=270,width=3in]{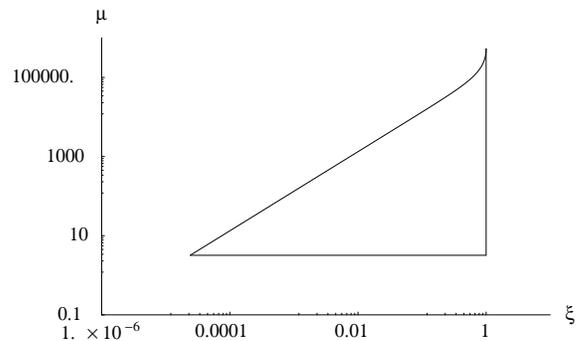}
    \caption{Bound on parameter space within EMBH model.}
    \label{limits}
\end{figure}

Condition $r_{dya}=r_{+}$ gives \cite{1998A&A...338L..87P}:
\begin{eqnarray}
    \mu=5.31\,10^5\frac{\xi}{\left(1+\sqrt{1-\xi^2}\right)^2}.
\label{bound}
\end{eqnarray}
It defines the maximal allowed mass for given charge of EMBH for which the electric field outside the horizon is overcritical one and can produce electron-positron pairs. The lower bound on EMBH mass comes from the maximal mass of the neutron star $M=3.2M_\odot$. Equality (\ref{bound}) can be inverted to get the lower bound on EMBH charge for this value, so finally we have the following ranges of basic parameters:
\begin{eqnarray}
    3.2=\mu_{min}< & \mu & <\mu_{max}=5.31\,\,10^5, \\
    2.41\,\,10^{-5}=\xi_{min}< & \xi & <\xi_{max}=1.
\end{eqnarray}
These bounds together with (\ref{bound}) are illustrated at fig. \ref{limits}.
\begin{figure}
    \centering
        \includegraphics[width=3in]{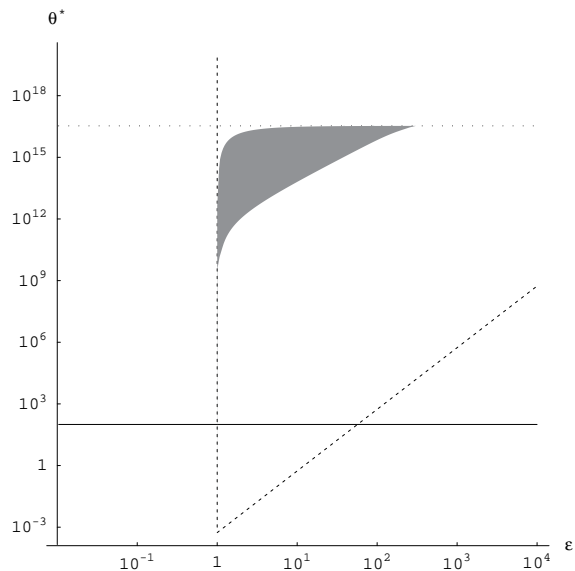}
    \caption{Cavallo and Rees diagram for EMBH model (details see in the text).}
    \label{theplim}
\end{figure}

Cavallo and Rees \cite{1978MNRAS.183..359C} suggested to describe the fireball with two parameters $\theta^{*}$ and $\epsilon$, where
\begin{eqnarray}
    \theta^{*}=\frac{E_0}{R_0^2}\frac{\sigma_T}{m_p c^2},
\end{eqnarray}
where $m_p$ is proton mass and $\epsilon$ is defined in (\ref{epsilon}). Parameter $\theta^{*}$ is tightly connected with optical depths with respect to electron-positron pairs $\tau_{pp}$ and electrons from the plasma admixture $\tau_{gas}$:
\begin{eqnarray}
    \theta^{*}=\tau_{pp}\frac{m_e}{m_p}\epsilon^3, \\
    \theta^{*}=\eta\tau_{gas}.
\end{eqnarray}

The diagram presented at fig. 1 of \cite{1978MNRAS.183..359C} allows to predict the evolution of the fireball. It is of great interest to find which restrictions on this picture are predicted by EMBH model. These constraints are shown in fig. \ref{theplim}.

The vertical dashed line denotes the pairs production threshold so to the right from this line pairs production occurs within the fireball. The inclined dashed line denotes the condition $\tau_{pp}=1$ and the fireball with parameters in the region above this curve is opaque due to pairs. The solid horizontal line denotes the condition $\tau_{gas}=1$ and the fireball is opaque due to plasma admixture above this line. The region of parameters allowed by EMBH model is represented by grey region. The upper boundary is given by the condition $\xi=\xi_{max}$ while the lower boundary corresponds to $\mu=\mu_{min}$. The maximal possible value of parameter $\theta^{*}=3.39\,\,10^{16}$ is shown by the dotted line. The maximal possible average particle energy produced by EMBH is $\epsilon=306$. At the same time, admissible initial conditions with small $\theta^{*}$ (up to $\theta^{*}=0$) lie in a very thin region close to the pairs production threshold $\epsilon=1$.

The values of parameter $\theta^{*}$ for EMBH case are much higher than ones considered by Cavallo and Rees. It implies that the energy is released in a much compact region of space and with much larger intensity. Clearly the fireball produced by EMBH cannot have $\epsilon\leq 1$, otherwise pairs would not be created. On this diagram the solid horizontal line corresponds to $B=10^{-2}$, the critical case for EMBH theory. For smaller values of $B$ this line moves up because it represents the condition $\theta^{*}=\eta=B^{-1}$. It can even cross the allowed region of parameters for the fireball for very small $B$. This means that below this line the fireball will be opaque only to pairs and not to the plasma, since the plasma admixture is tiny.

From fig. \ref{theplim} it can be seen that the fireball produced by EMBH is almost always opaque both to pairs and to plasma, i.e. its initial conditions lay in the the region III defined in \cite{1978MNRAS.183..359C}. This implies that the fireball necessary undergoes adiabatic expansion with $\theta^{*}\propto\epsilon^3$. Therefore, cases I and IV described by Cavallo and Rees are irrelevant for EMBH model. The case II is only partially relevant for almost pure fireballs when the plasma admixture is very small $B\ll 10^{-10}$.

\section{Conclusions}

We compared existing isotropic models of GRBs, so called fireball models. It is shown that the crucial difference between our approach and other models in the literature is the presence of the rate equation which accounts for electron-positron pairs densities evolution during expansion of the fireball. This results in quantitative difference between predictions of our quasi-analytic model and analytic models in the literature. Considering its significance we conclude that in order to compare theory and observations it is necessary to take into account rate equation together with energy and mass conservation conditions.

Another important difference is the presence of bound on baryonic loading parameter $B_c=10^{-2}$ which comes from violation of constant thickness approximation used in our quasi-analytic model. The same bound should be present in all analytic models in the literature. As a consequence, the broadening of the relativistic shell resulting from the fireball never happens before it reaches transparency. Besides, the gamma factor does not reach saturation and the value $\gamma=\eta=B^{-1}$ is only asymptotic one for $B<10^{-2}$.

Bounds on dimensionless quantities describing initial conditions for the fireball are deduced from EMBH theory. It is shown that the fireball appearing as the result of energy extraction process from electro-magnetic black hole is always opaque both to its electron-positron pairs and electons of the baryonic material which the fireball absorbs during early stages of expansion. As a consequence the fireball resulting from EMBH necessarly undergoes adiabatic expansion all the way to transparency.

\bibliographystyle{aip}
\bibliography{baryonic}

\end{document}